\documentstyle [prl,aps,twocolumn,psfig]{revtex}
\newskip\humongous \humongous=0pt plus 1000pt minus 1000pt

\newif\ifdtup

\relax

\begin{document}

\newcommand{\newc}{\newcommand}

\newc{\be}{\begin{equation}}
\newc{\ee}{\end{equation}}
\newc{\ba}{\begin{eqnarray}}
\newc{\ea}{\end{eqnarray}}
\newc{\bea}{\begin{eqnarray}}
\newc{\eea}{\end{eqnarray}}
\newc{\D}{\partial}
\newc{\ie}{{\it i.e.} }
\newc{\eg}{{\it e.g.} }
\newc{\etc}{{\it etc.} }
\newc{\etal}{{\it et al.}} 

\newc{\ra}{\rightarrow}
\newc{\lra}{\leftrightarrow}
\newc{\no}{Nielsen-Olesen }
\newc{\tp}{'t Hooft-Polyakov }
\newc{\lsim}{\buildrel{<}\over{\sim}}
\newc{\gsim}{\buildrel{>}\over{\sim}}
\title{Searching for Long Strings in CMB Maps}

\bigskip

\author{L. Perivolaropoulos}
  
\address{Department of Physics,
University of Crete\\ 
P.O.Box 2208, 710 03 Heraklion, Crete; Greece\\
e-mail: leandros@physics.uch.gr}

\date{\today}
\maketitle

\begin{abstract}
Using analytical methods and Monte Carlo simulations, we analyze new statistics 
designed to detect isolated step-like discontinuities which are coherent over 
large areas of Cosmic Microwave Background (CMB) pixel maps.
Such coherent temperature discontinuities are predicted by the {\it 
Kaiser-Stebbins} effect to form due to long cosmic strings present in our 
present horizon. The background of the coherent step-like seed is assumed to be 
a scale invariant Gaussian random field which could have been produced by a 
superposition of seeds on smaller scales and/or by inflationary quantum 
fluctuations. 
We find that the proposed statistics can  detect the 
presense of a coherent discontinuity at a sensitivity level almost an order of magnitude better compared to more conventional statistics 
like the skewness or the kurtosis.

\end{abstract}

\narrowtext


\section{Introduction}
\noindent

The major progress achieved during the past 15 years in both theory and  
cosmological observations has turned the search for the origin of
cosmic structure into one of the most exciting fields of scientific 
research. 
Despite the severe constraints imposed by detailed observational
data on theories for structure formation the central question remains open:
{\it What is
the origin of primordial fluctuations that gave rise to structure in the 
universe?} 
 Two classes of theories attempting to answer this question have emerged during 
the past twenty years
and have managed to survive through the observational constraints with only 
minor adjustments.
 
According to the first class, primordial fluctuations are produced by quantum 
fluctuations of a
linearly coupled scalar field during a period of inflation 
\cite{gp82}. These
fluctuations are subsequently expected to become classical and provide the 
progenitors of structure
in the universe. Because of the extremely small linear coupling of the scalar 
field, needed to
preserve the observed large scale homogeneity, the inflationary perturbations 
are expected by the
central limit theorem, to obey Gaussian statistics. This is not the case for the 
second class of
theories.
 
According to the second class of theories 
\cite{k76},
primordial perturbations are provided by {\it seeds} of trapped energy density 
produced during
symmetry breaking phase transitions in the early universe. Such symmetry 
breaking is predicted by
Grand Unified Theories (GUT's) to occur at early times as the universe cools and 
expands. The
geometry of the produced seeds, known as {\it topological defects} is determined 
by the topology of
the vaccuum manifold of the physically realized GUT. Thus the defects may be 
pointlike (monopoles),
linelike (cosmic strings), planar (domain walls) or collapsing pointlike 
(textures).
 
The cosmic string theory \cite{v81} for structure formation is the oldest 
and (together with
textures \cite{t89}) best studied theory of the topological defect class. By 
fixing its single free
parameter $G\mu$ ($\mu$ is the {\it effective} mass per unit length of the 
wiggly string $G$ is
Newtons constant and we have used units with $c=1$) to a value consistent with 
microphysical requirements coming from GUT's ($G\mu \simeq 10^{-6}$), the 
theory is consistent with the noise in pulsar signal arrival times assuming that the 
noise is due to gravitational radiation emitted by the defect network 
\cite{cbs96}. 
It may automatically account for large scale filaments and sheets 
\cite{v86}, galaxy
formation at epochs $z\sim 2-3$ \cite{bk87} and galactic 
magnetic fields
\cite{v92b}. It
can also provide large scale peculiar velocities \cite{v92a} and is consistent with the amplitude, 
spectral index
\cite{bbs88}   and the statistics 
\cite{g90}  
 of the cosmic microwave background (CMB) anisotropies measured by the COBE 
collaboration
\cite{s92} on large angular scales 
($\theta\sim 10^\circ$). Other planned CMB experiments \cite{ss95}
of equally high quality but on smaller angular scales are expected to provide a 
wealth of information within the next few years.
 
The CMB observations provide a valuable direct probe for identifying signatures 
of cosmic strings.
The main mechanism by which strings can produce CMB fluctuations on angular
scales larger than 1-2 degrees 
has been well studied both analytically 
\cite{vs90}  and
using numerical simulations \cite{bbs88} and is known as the {\it
Kaiser-Stebbins effect} \cite{ks84}.    According to this
effect, moving long strings present between the time of recombination $t_{rec}$ 
and the present time
$t_0$, produce step-like temperature discontinuities between photons that reach 
the observer through
opposite sides of the string. These discontinuities are due to the peculiar 
nature of the spacetime
around a long string which even though is {\it locally} flat, {\it globally} has 
the geometry of
a cone with deficit angle $8\pi G\mu$. The magnitude of the discontinuity is 
proportional to the
deficit angle, to the string velocity $v_s$ and depends on the relative 
orientation between the unit
vector along the string ${\hat s}$ and the unit photon wave-vector ${\hat k}$. 
It is given by
\cite{vs90}  
\begin{equation} 
{{\delta T}\over T}=\pm 4\pi G\mu v_s \gamma_s {\hat k} \cdot ({\hat v_s}\times 
{\hat s})
\end{equation}
where $\gamma_s$ is the relativistic Lorentz factor and the sign changes when 
the string is
crossed. The angular scale over which this discontinuity persists is given by 
the radius of
curvature of the string which according to simulations 
\cite{bb88} is approximately equal
to the horizon scale. 
The growth of the horizon from $t_{rec}$ to $t_0$ results in a superposition
of a large number of step-like temperature seeds of all sizes starting from
about $2^\circ$ (the angular size of the horizon at $t_{rec}$) to about 
$180^\circ$ (the present horizon scale). By the central limit theorem this
large number of superposed seeds results in a pattern of fluctuations that
obeys Gaussian statistics. Thus the probability distribution for the 
temperature of each pixel of a CMB map with resolution larger than about
$1^0 - 2^0$ is a Gaussian  \cite{ac96}. 
It has therefore been considered to be impossible
to distinguish structure formation models based on cosmic strings from 
corresponding models based on inflation, using CMB maps with resolution 
angle larger than $1^0 -2^0$. 
Theoretical studies have therefore focused
on identifying the statistical signatures of cosmic strings on angular scales
less than $1^0$ \cite{g90}  where the number of superposed seeds is smaller and therefore
the non-Gaussian character of fluctuations is expected to be stronger 
\footnote{The non-Gaussian features for texture maps are stronger than thosed
of cosmic strings mainly because of the generically smaller number of
textures per horizon volume \cite{g96}.}

These efforts however  have been faced with the complicated and model
dependent physical processes occuring on small angular scales. Such 
effects include isolated foreground point sources, recombination physics,
string properties on small scales (kinks, loops etc) which require
detailed simulations of both the string network and the cosmic background,
in order to be properly taken into account.  Even though there are preliminary
efforts for such detailed simulations
\cite{ac96},  it has become clear
that it will take some time before theory and experiments on angular
scales less than a few arcmin reach accuracy levels leading to detectable
non-Gaussian string signatures.

 An alternative approach to the problem is instead of focusing on small
scales where the number of superposed seeds is small, to focus on larger
angular scales where despite the large number of superposed seeds there 
is also coherence of induced fluctuations on large angular scales. Fluctuations
on these scales may be viewed as a superposition of a Gaussian scale 
invariant background coming mainly from small scale seeds plus a small 
number of step-like discontinuities which are coherent and persist on angular 
scales of order $100^0$. These are produced by long strings present in
our present horizon. {\it Our goal is to find a statistic optimized to detect 
this large scale coherence and use it to find the minimum amplitude of a coherent discontinuity that can be detected at the $1\sigma-2\sigma$ 
level relatively to a given scale 
invariant or noise dominated Gaussian background.}
Such a statistic is equally effective on {\it any} angular resolution scale and 
its effectiveness is only diminished as the number of pixels of the CMB map is 
reduced or the noise is increased. 
The statistical variable we focus on, in sections 2 and 3 is the Sample Mean 
Difference (SMD) of temperatures between large neighbouring sectors of a CMB map. These sectors 
are separated by a random straight line in two dimensional maps or by a random 
point  in one dimensional maps. The union of the two sectors gives back the 
complete map. 
We show that the statistics of the SMD variable are much more sensitive in 
detecting the presence of a coherent step-like seed than conventional statistics 
like the skewness or the kurtosis. 

We also discuss an 
alternative statistic, the Maximum Sample Difference that is more sensitive in certain cases but less 
robust than the SMD. This statistic is based on finding the 
maximum from a large sample of temperature differences between large 
neighbouring sectors of CMB maps. 
We show that for noise dominated data the MSD statistic is even more 
sensitive than the SMD statistic. However, this sensitivity gets 
rapidly reduced when significant correlations are introduced in the underlying 
Gaussian data. Thus the MSD statistic is more sensitive but less robust 
compared to the SMD statistic.  

The structure of this paper is the following: 
In the next section titled 'Sample Mean Difference' we study analytically the 
statistics of the SMD variable and show that its average value is a sensitive 
quantity in detecting the presence of a randomly positioned step-function on top 
of a Gaussian map. We then compare with the sensitivity of the statistics {\it 
skewness} and {\it kurtosis}. We find that the sensitivity of the SMD statistic 
is significantly superior to that of skewness and kurtosis in detecting the step 
function. These analytical results are shown for the case of one-dimensional 
maps but the extension to the case of two dimensional maps is straightforward.

In the third section titled 'Monte Carlo Simulations' we perform Monte Carlo 
simulations of Gaussian maps with 
and without step-like discontinuities in one and two dimensions. Applying the statistics 
skewness, kurtosis and average of SMD on these maps we verify the analytical results of 
section 2 and find the minimum step-function amplitude that is detectable by the 
average SMD statistic. In section 4 we show both analytically and numerically 
that the MSD statistic can be significantly more sensitive and accurate even 
compared to the SMD statistic when applied to noise dominated data.

Finally is section 5 we conclude, summarise and discuss 
the prospect of applying the average of SMD statistic and the MSD statistic 
to presently available CMB 
maps including the COBE  data. That analysis is currently in progress 
\cite{ps97}, and will be presented separately.

\section{\bf Sample Mean Difference}

Consider an one dimensional array of $n$ pixel variables $x_n$. Let these 
variables be initially distributed according to a standardised Gaussian 
probability distribution. Consider now a step-function of amplitude $2\alpha$  
superposed so that the discontinuity is between pixels $i_0$ and $i_0 + 1$ (Fig. 
1). The new probability distribution for a random pixel variable $x$ is 

\be
P(x) = {f \over {\sqrt{2\pi}}} e^{-{{(x-\alpha)^2} \over 2}} + 
{{1-f} \over {\sqrt{2\pi}}} e^{-{{(x+\alpha)^2} \over 2}}
\ee
where $f = {{i_0} \over n}$. We are looking for a statistic that will optimally 
distinguish between a Gaussian array with a superposed step-function and a 
Gaussian array without one. The obvious statistics to try first are the moments 
of the distribution (2) with $\alpha = 0$ and $\alpha \neq 0$. \footnote{With no loss of generality we may assume $\alpha$ positive or $0$ 
because from the statistical point of view there is $\alpha \leftrightarrow -\alpha$ symmetry.}
\begin{figure}
\centerline{
\psfig{figure=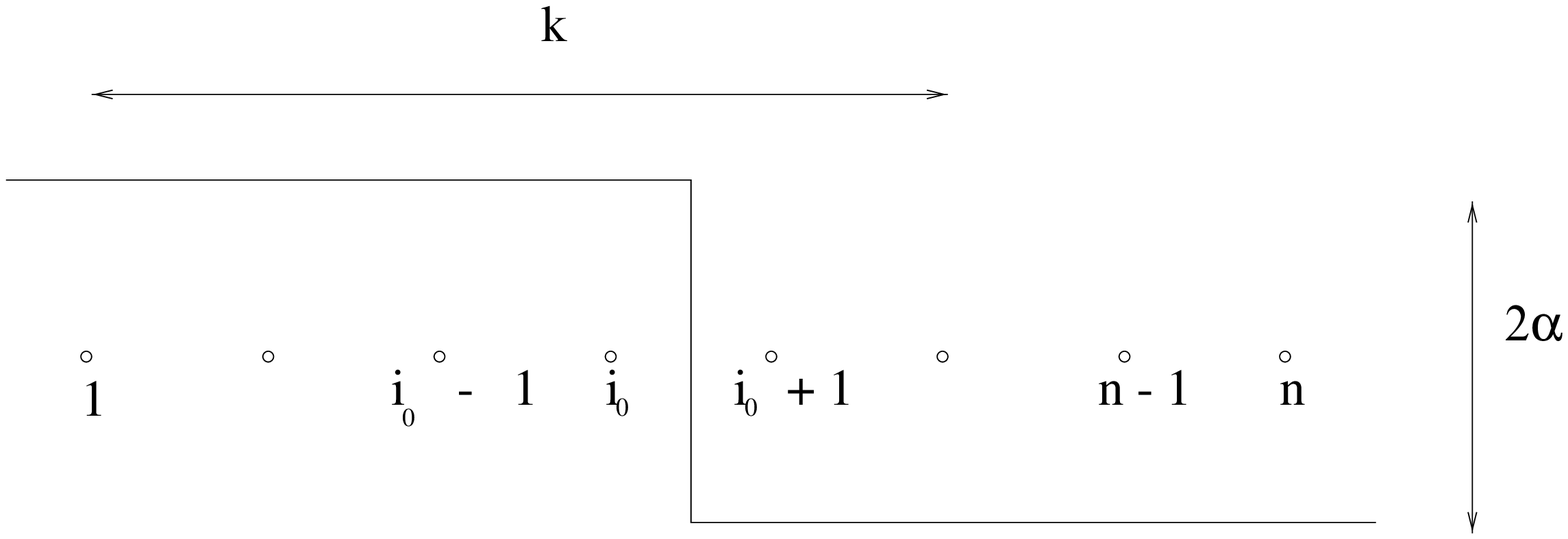,height=2.0in,angle=0}
\psdraft
}
Figure 1: A large scale coherent step-function discontinuity superposed on a one 
dimensional pixel map.

\end{figure}
The moment generating function corresponding to (2) is:
\be
M(t) = f e^{\alpha t + {t^2 \over 2}} +
(1-f) e^{-\alpha t + {t^2 \over 2}}
\ee
The mean $\mu (\alpha, f)$, variance $\sigma^2 (\alpha,f)$, skewness 
$s(\alpha,f)$ and kurtosis $k(\alpha,f)$ can be obtained in a straightforward 
way by proper differentiation of $M(t)$ as follows:
\ba
\mu (\alpha, f)&\equiv & <X> = \alpha f - \alpha (1-f) \nonumber \\
\sigma^2 (\alpha,f) &\equiv & < (x - \mu )^2 > = 
1 + 4 \alpha^2 f (1 - f) \nonumber \\
s(\alpha,f) &\equiv & {{< (x - \mu )^3 >}\over \sigma^3} =
{{8 \alpha^3 f (1-3 f + 2 f^2)} \over {(1 + 4 \alpha^2 f (1 - f))^{3/2}}} \nonumber
\ea
\begin{eqnarray*}
&k&(\alpha,f) \equiv   {{< (x - \mu )^4 >}\over \sigma^4}\\ 
&=&{{3 + 8 \alpha^2 f (3 + 2\alpha^2 -3f^2 - 8 \alpha^2 f + 12 \alpha^2 f^2 
-6\alpha^2 f^3)} \over {(1 + 4 \alpha^2 f (1 - f))^4}}
\end{eqnarray*}

For $\alpha = 0$ we obtain the Gaussian values for the skewness and the kurtosis 
$s(0,f) = 0$ , $k(0,f) = 3$ as expected. For $\alpha \neq 0$ the moments deviate 
from the Gaussian values. In order to find the minimum value of $\alpha$ for 
which the moments can distinguish between a Gaussian pattern and a Gaussian+Step 
pattern we must compare the deviation of moments from their Gaussian values with 
the standard deviation of the sample moments.
The mean values of the skewness and the kurtosis are easily obtained by 
integrating with respect to $f$ from 0 to 1 i.e. assuming that it is equally 
probable for the step-function to be superposed at any point of the lattice.
\ba
{\bar s}(\alpha) & = &<s(\alpha, f)> = \int_0^1 df s(\alpha,f) = 0 \\
{\bar k}(\alpha) & = &<k(\alpha, f)> = \int_0^1 df k(\alpha,f)
\ea
These values are to be compared with the standard deviations of the moments, 
obtained as follows:
The variance of the skewness over several n-pixel array realizations with fixed 
$f$ and $\alpha$ is
\be
\Delta s^2 (\alpha,f) = <({\hat s} - s)^2 >
\ee
where ${\hat s} \equiv {{s_1 + ... + s_n} \over n}$ is the sample skewness from 
a given pixel array realization, $s$ is the actual skewness (for $n \rightarrow \infty$) and $s_i \equiv 
{{(x_i - \mu)^3} \over \sigma^3}$. Now
\be
<{\hat s}> = {{n < s_1>} \over n} = <s_1> = s
\ee
Also
\be
<{\hat s}^2> = {1\over n} <s_j^2> + (1-{1\over n}) <s_j>^2
\ee
where $j$ {\it any} pixel number ($j \in [1,n]$).
Thus
\be
\Delta s^2 (\alpha,f) = {1\over n} (<s_j^2> - <s_j>^2) 
\ee
Similarly for the variance of the sample kurtosis we have
\be
\Delta k^2 (\alpha,f) = {1\over n} (<k_j^2> - <k_j>^2)  
\ee
with $k_j=
{1 \over \sigma^4}  (x_j - \mu)^4
$
and  
$  
<k_j^2> = {1 \over \sigma^8} <(x_j-\mu)^8>
$.
It is straightforward to obtain all the above moments by differentiating
the generating functional and using
\be
< x_j^n > ={{d^n M} \over {d t^n}} \vert_{t=0} 
\ee
Now the minimum value $\alpha_{min}$ of $\alpha$ detectable at $1\sigma$
level is obtained from the equations
\ba
\int_0^1 df [s(\alpha_{min},f) - \Delta s(\alpha_{min},f)] &=& 0 \\
\int_0^1 df [(k(\alpha_{min},f)-3) - \Delta k(\alpha_{min},f)] &=& 0  
\ea
Since (from eq. (4)) ${\bar s} (\alpha) = 0$ which is equal to the 
Gaussian value, the skewness can only be used to detect a step function
by comparing the standard deviation $\Delta{\bar s}$ for $\alpha = 0$ and
$\alpha \neq 0$. By demanding $\Delta {\bar s} (\alpha_{min}) \leq 2 
\Delta {\bar s} (\alpha = 0)$ we obtain $\alpha_{min} \leq 2.5 $. This
result is independent of the number of pixels $n$. For the kurtosis
we obtain from eqs. (10, 13) $\alpha_{min}  \simeq 4 $ for $n = 10^3$ while
for $ \alpha_{min}  =0.5$, $n\simeq 10^6$ is required.

Using the alternative test i.e. demanding $\Delta {\bar k} (\alpha_{min}) 
\geq 2 \Delta {\bar k} (\alpha = 0) $ we obtain $ \alpha  \geq 2$ and this
result is independent of the number of pixels $n$ as in the case of skewness.
Thus for the usual pixel maps where $n$ is up to $O(1000)$ the kurtosis is
not able to detect a step function with $| \alpha | \leq 2$ at the $1\sigma$
level. \footnote{As in all cases discussed in this paper $\alpha$ is measured
in units of standard deviation (rms) of the underlying Gaussian map.}
This result remains unchanged for other statistical variables defined by 
{\it local} linear combinations of pixels (e.g. differences of neighbouring 
pixel variables \cite{g90}) 
since the effect of a single discontinuity remains negligible if the long range 
coherence is not taken into account.

For CMB temperature maps with $({{\delta T} \over T})_{rms} \simeq 
2 \times 10^{-5}$ the detectable value of $G\mu$ is
\ba
\alpha &\equiv & 4\pi G \mu (v_s \gamma_s)\cos \theta > 4 \times 10^{-5} \nonumber \\
& \Rightarrow & G \mu (v_s \gamma_s)\cos \theta \gsim  4 \times 10^{-6}
\ea
where $\theta$ is an angle obtained from the relative orientation of the string 
with respect to the observer.
According to simulations $<v_s \gamma_s >_{rms} \simeq 0.2 $ and for
$G\mu < 2 \times 10^{-5}$ the detection of the Kaiser-Stebbins effect 
using statistics based on skewness and kurtosis is not possible. This excluded range 
however includes all the cosmologically interesting 
values of $G\mu$.

It is therefore important to look for alternative statistical
variables that are more sensitive in detecting the presence
of coherent discontinuities superposed on Gaussian maps. It will be
shown that the Sample Mean Difference (SMD) is such a statistical variable. A similar 
and even more effective statistic, the Maximum Sample Difference will be discussed in section IV.

Consider a pixel array (Fig. 1) of $n$ pixel Gaussian random variables
$X_j$ with a step function covering the whole array, superposed such 
that the discontinuity is located just after pixel $i_0$. To every pixel
$k$ of the array we may associate the random variable $Y_k$ defined as
the difference between the mean value of the pixels 1 through $k$ minus the 
mean value of the pixels $k+1$ through $n$.
It is straightforward to show that
\ba
Y_k &=& \Delta {\bar X}_k + 2 \alpha {{n-i_0} \over {n-k}} \hspace{1cm}
k\in [1,i_0] \\
Y_k  &=& \Delta {\bar X}_k + 2 \alpha {i_0 \over k} \hspace{1cm}
k\in [i_0 ,n-1] 
\ea
where $\Delta {\bar X}_k  \equiv {1\over k} \sum_{j=1}^k X_j - {1 \over {n-k}}
\sum_{j=k+1}^n X_j$. Thus we have constructed a new array $Y_k$, $(k=1,...,n-1)$ 
from the sample mean differences (SMD) of the original array. We will focus on 
the average value $Z$ of the SMD defined as:
\be
Z={1\over {n-1}} \sum_{k=1}^{n-1} Y_k
\ee
Using eqs. (15-17) we obtain
\be
Z={1\over {n-1}} [\sum_{k=1}^{n-1} \Delta{\bar X}_k + 2\alpha 
(\sum_{k=1}^{i_0} {{1-i_0 / n} \over {1-k/n}} + 
\sum_{k=i_0 + 1}^{n-1} {{i_0/n} \over {k/n}}) ]
\ee
With the definitions $f\equiv {i_0 / n}$ and $\xi \equiv {k/n}$ and the
assumption $n>>1$ we obtain:
\be
Z = {1 \over {n-1}} \sum_{k=1}^{n-1} \Delta {\bar X}_k - 2\alpha 
[(1-f) \ln(1-f) + f \ln f]
\ee
Thus the mean of $Z$ over many realizations of the array is
\be
<Z> = {1 \over {n-1}} \sum_{k=1}^{n-1} <\Delta {\bar X}_k>- 4\alpha 
[\int_0^1 df \hspace{1mm} f \ln f] = \alpha
\ee
The variance of $Z$ is due both to the underlying Gaussian map and to the 
variation of $f=i_0 /n $ (assuming $\alpha$ fixed). The variance due to the  
gaussian background is
\be
\sigma_{1,Z}^2 = {1\over {(n-1)^2}} \sum_{k=1}^{n-1} ({1\over k} + 
{1\over {n-k}}) \simeq \epsilon \int_\epsilon^{1-\epsilon} {{d\xi} \over {\xi 
(1-\xi)}}
\ee
where $\epsilon = O({1\over n})$, $\xi = k/n$, $n>>1$ and we have used the fact 
that the variance of the sample mean of a standardized Gaussian population with 
size $j$ is  $1\over j$. Now from eq. (21) we obtain
\be
\sigma_{1,Z}^2 \simeq -\epsilon \ln \epsilon^2 \simeq {{2 \ln n}\over n}
\ee
The variance of the $f$-dependent part of Z is
\be
\sigma_{2,Z}^2 = <Z_2^2> - <Z_2>^2
\ee
where $Z_2 \equiv -2\alpha [(1-f) \ln(1-f) + f \ln f] $. From eq. (20) we have 
$<Z_2> = \alpha$ and $<Z_2^2>$ is easily obtained as
\be
<Z_2^2> = \int_0^1 df \hspace{1mm} Z_2^2 (f) \simeq {4\over 3} \alpha^2
\ee
Thus
\be
\sigma_Z^2 \equiv \sigma_{1,Z}^2 + \sigma_{2,Z}^2 = {{2\ln n} \over n} + {1\over 
3}\alpha^2
\ee
In order to be able to distinguish between a Gaussian+Step map and a purely 
Gaussian one, at the $m \sigma$ level we demand that
\be
|Z_{\alpha \neq 0} - Z_{\alpha = 0}| \geq m \hspace{0.2cm} \sigma_{1Z}
\ee
where we have used the variance $\sigma_{1Z}$ of a given realization.
This implies that the minimum value of $\alpha$, $\alpha_{min}$ that can be 
detected using this test is
\be
\alpha_{min} = m({{2\ln n}\over n})^{1/2}
\ee
and for $n=O(10^3)$ we obtain $\alpha_{min} \simeq m \hspace{0.2cm} 0.2$ which for $m=1$ is about {\it an order 
of magnitude improvement} over the corresponding sensitivity of tests based on 
the moments skewness and kurtosis. The reason for this significant improvement 
is the fact that the SMD statistical variable picks up the {\it coherence} 
properties introduced by the step function on the Gaussian map. The moments on 
the other hand pick up only local properties of the pixels and do not amplify 
the long range coherence of the step-like discontinuity.

Our analysis so far has assumed that the Gaussian variables $X_j$ are 
independent and that the only correlation is introduced by the superposed 
step-function. In a realistic setup however the underlying Gaussian map will be 
scale invariant and thus there will be correlations among the pixels. These 
correlations will also be affected by the instrument noise. In addition, our 
analysis has been limited so far to one dimensional maps while most CMB 
experiments are now obtaining two-dimensional maps. In order to take all these 
effects into account we need to apply the statistics of the SMD variable onto 
maps constructed by Monte Carlo simulations. This is the focus of the following 
section. 

\section{\bf Monte-Carlo Simulations}
We start by constructing an array of $n$ Gaussian random variables $X_j$, 
$j=1,...,n$ with a power spectrum $P(k)=k^{-m}$. Thus the values $X_j$ associated 
with the pixel $j$ are obtained as the Fourier transform of a function $g(k)$ 
($k=1,...,n$) with the following properties:
\begin{itemize}
\item
For each $k$, the amplitude $|g(k)|$ is an independent Gaussian random variable with 0 
mean and variance $P(k) = 1/k^m$.
\item
The phase $\theta_k$ of each Fourier component $g(k)$ is an independednt random 
variable in the range $[0,2\pi]$ with uniform probability distribution 
$P(\theta_k) = {1\over {2 \pi}}$.
\item
The Fourier components are related by complex conjugation relations neeeded to 
give a {\it real} variable $X_j$.
\end{itemize}
The discrete Fourier transform definition used is
\be
X_j = {1\over \sqrt{n}} \sum_{k=1}^n g(k) e^{2\pi i (k-1)(j-1)/n}
\ee
and the numerical programming was implemented using {\it Mathematica}
\cite{w91}.
The array $X_j$ obtained in the way described above is then 
standardized to the array $X_j^s$, with 
\be
X_j^s \equiv {{(X_j - \mu)}\over {\sigma}}
\ee
where $\mu$ and $\sigma^2$ are the sample mean and sample variance for the 
realization of the array $X_j$. A new array $X_j^\prime$ is then constructed by 
superposing to the array $X_j^s$ a step function of amplitude $2\alpha$ with 
discontinuity at a random point $i_0$. The array $X_j^\prime$ is thus obtained 
as
\be
X_j^\prime = X_j^s + \alpha {{j-i_0} \over {|j-i_0|}}, \hspace{1cm} j=1,...,n
\ee
Next we apply the statistics discussed in the previous section to several 
realizations of the arrays $X_j^s$ and $X^\prime$ in an effort to find the most 
sensitive statistic that can distinguish among them. Our goal is to also find 
the minimum value of $\alpha $ that can be distinguished by that statistic at the 
$1\sigma$ level, thus testing the analytical results of the previous section.

We have used a lattice with 2000 pixels and a scale invariant power spectrum 
which for one-dimesional data is $P(k)=k^{-1}$.
In Table 1 we show the results for the skewness, the kurtosis and the average 
SMD for the $X_j$ arrays, with $\alpha =$0, 0.25, 0.50 and 1.0. The SMD average 
was obtained as in section 2 by first constructing the array of sample mean 
differences and then obtaining its average value, predicted to be equal to 
$\alpha$ by the analytical study of section 2.

These statistics were applied to 50 random realizations of the array $X_j^s$. 
The mean values of the statistics considered with their $1\sigma$ standard 
deviations obtained over these 50 realizations are shown in the following Table 
1.

{\bf Table 1}: A comparison of the effectiveness of the statistics considered, in 
detecting the presence of a coherent step discontinuity with amplitude $2\alpha$ 
relative to the standard deviation of the underlying scale invariant Gaussian map. 
\vskip 0.1cm
\begin{tabular}{|c|c|c|c|}\hline
{\bf $\alpha $ }&{\bf  Skewness }&{\bf  Kurtosis }&|{\bf SMD Average }| \\ \hline
0.00 &$0.01 \pm 0.11$     &$2.97 \pm 0.19$     &$0.02 \pm 0.31$\\ \hline
0.25 &$0.01 \pm 0.11$     &$2.95 \pm 0.20$     &$0.25 \pm 0.33$ \\ \hline
0.50 &$0.02 \pm 0.11$     &$2.88 \pm 0.21$     &$0.48 \pm 0.38$ \\ \hline
1.00 &$0.03 \pm 0.20$     &$2.82 \pm 0.32$     &$0.98 \pm 0.48$ \\ \hline
\end{tabular}
\vspace{3mm}

The analytical prediction of section 2 for the SMD average value $\alpha$ is in 
good agreement with the results of the Monte Carlo simulations. The standard 
deviation of this result is not in such a good agreemnent with the analytical 
prediction because the assumption of complete independence among pixels made by 
the analytical treatment is not realized in the Monte Carlo simulations. Ôhere a 
scale invariant spectrum was considered and thus there was a non-trivial 
correlation among the pixels of the arrays.

A simple way to further improve the sensitivity 
of the SMD statistical variable is to ignore a number $l$ of boundary pixels of 
the SMD array, thus constructing its average using the Sample Mean Differences 
of pixels $l+1,...,n-l$. From eq (21), the variance of the SMD  for these pixels 
is significantly lower than the corresponding variance of the $2l$ pixels close 
to the boundaries. In addition, if the step is located within the central $n-2l$ 
pixels the SMD average may be shown to be larger than $\alpha$ thus further 
amplifying the step signature. For $l=150$ the variance of the SMD average {\it 
is reduced} by about $20$\% while the SMD average is {\it increased} by 
about $20$\% thus allowing the detection of steps as low as $\alpha = 0.25$ at 
the $1\sigma$ level. The price to pay for this sensitivity improvement is the 
reduction of the effective pixel area where the search for steps is made.

We have also used the SMD statistical variable for non-scale invariant power 
spectra and found that it works better for $P(k) = k^{-m}$ with $0 \leq m <1$ 
than for $m>1$. This is to be expected because large values of $m$ imply larger 
correlations among pixels which in turn leads to a smaller number of effectively 
independent pixels and thus a larger value for the variance of the SMD average. 

It is straightforward to generalize the one dimensional Monte Carlo simulations 
to two dimensions. In that case we use the two-dimensional discrete Fourier 
transform as an approximation to an expansion to spherical harmonics. This 
approximation is good for small area maps of the celestial sphere. We used the 
following definition of the two dimensional discrete Fourier transform.
\be
X(i,j) = {1 \over n} \sum_{k_1, k_2 = 1}^n  g(k_1,k_2) e^{2\pi 
i[(i-1) (k_1 -1) +(j-1) (k_2 - 1)]/n}
\ee
refering to a $n \times n$ square lattice. In order to construct the background 
of scale invariant Gaussian fluctuations we used $g(k_1,k_2)$ as a complex 
random variable. For scale invariance, the amplitude of $g(k_1,k_2)$ was 
obtained from a Gaussian probability distribution with 0 mean and variance
\be
\sigma^2 (k_1,k_2) = P(k_1,k_2) = {1\over {k_1^2 + k_2^2}}
\ee

The corresponding phase $\theta_{k_1,k_2}$ for the $(k_1,k_2)$ mode was also 
determined randomly from a uniform probability distribution 
$P(\theta_{k_1,k_2})={1\over {2\pi}}$ in order to secure Gaussianity for the map 
$X(i,j)$. 

The corresponding map with a superposed coherent step discontinuity 
was obtained from the standardized Gaussian map $X^s (i,j)$ as
\be
X^\prime (i,j) = X^s(i,j) + \alpha {{j-a \hspace{1mm} i - b} \over
{|j-a \hspace{1mm} i - b|}}
\ee
where 
\ba
a&=&{{y_2-y_1} \over {x_2 - x_1}}\\
b &=& y_1 - a \hspace{1mm} x_1
\ea
i.e. the line of step discontinuity $j=a \hspace{1mm} i +b$ is determined by the 
two random points $(x_1,y_1)$ and $(x_2,y_2)$ of the map $X(i,j)$. The skewness and kurtosis of the two maps are obtained in the usual way. 
For example for the standardized Gaussian 
map $X^s (i,j)$ we have
\ba
s &=& {1\over n^2} \sum_{i,j}^n X^s (i,j)^3 \\ 
k &=& {1\over n^2} \sum_{i,j}^n X^s (i,j)^4
\ea

The SMD statistical variable is obtained by considering a set of random 
straight lines bisecting the map and for each line taking the difference of the 
sample means from the two parts of the map. For example consider a line defined 
by the random points $(x_1,y_1)$ and $(x_2,y_2)$ of the map. The line equation 
is $j=a \hspace{1mm} i + b$ with $a$, $b$ obtained from eqs. (34) and (35). The 
SMD obtained from this line is
\be
{SMD} = {S_1 \over n_1} - {S_2 \over n_2}
\ee
where
\ba
S_1 &=& \sum_{i=1}^n \sum_{j=Max[(a \hspace{1mm} i + b),1]}^n X^s(i,j) \\     
S_2 &=& \sum_{i=1}^n \sum_{j=1}^{Min[(a \hspace{1mm} i + b),n]} 
X^s(i,j)    
\ea 
and  $n_1$, $n_2$ are the corresponding numbers of terms in the sums. For a 
Step+Gaussian map, the index $^s$ get replaced by $^\prime$. 

The average and variance of the SMD  
is obtained by averaging over a large number 
of random test lines $(a,b)$ and a large number of map realizations. The results 
of the application of the three statistics (skewness, kurtosis and SMD average) 
on $30 \times 30$ scale invariant Gaussian maps for various values of step 
amplitudes $\alpha $ are shown in Table 2. Uncorrelated Gaussian noise was also 
superposed on the signal with 
signal to noise ratio $2.0$. The
random points defining the test lines were excluded from the outermost three 
rows and columns of the maps thus reducing somewhat the 
variance of the SMD average.      

{\bf Table 2}: A comparison of the effectiveness of the statistics considered in 
two dimensional maps. The signal to noise ratio was $2.0$. Points defining the 
line discontinuities were excluded from the three outermost rows and columns of the maps.
\vskip 0.1cm
\begin{tabular}{|c|c|c|c|}\hline
{\bf $\alpha $ }&{\bf  Skewness }&{\bf  Kurtosis }&|{\bf SMD Average }|\\ \hline
0.00 &$0.04 \pm 0.13$     &$3.00 \pm 0.20$     &$0.01 \pm 0.03$\\ \hline
0.25 &$0.02 \pm 0.08$     &$2.97 \pm 0.13$     &$0.14 \pm 0.09$ \\ \hline
0.50 &$0.05 \pm 0.14$     &$2.91 \pm 0.24$     &$0.34 \pm 0.19$ \\ \hline
1.00 &$0.02 \pm 0.24$     &$2.95 \pm 0.30$     &$0.56 \pm 0.31$ \\ \hline
\end{tabular}
\vspace{3mm}

The results of Table 4 are in qualitative agreement with those of Table 1 and 
with the analytical results valid for the one dimensional maps. 
Clearly the details of the one dimensional analysis are not valid in the two 
dimensional case and so the agreement can not be quantitative.
The results still indicate however that the SMD statistic is significantly more 
sensitive compared to conventional statistics for the detection of coherent 
discontinuities on CMB maps. This statistic can detect coherent discontinuities 
with minimum amplitude $\alpha_{min} \simeq 0.5$ at the $1\sigma$ to $2\sigma$ level where 
$\alpha$ is the amplitude relative to the standard deviation of the underlying 
scale invariant Gaussian map.

\section{Maximum Sample Difference}

An alternative statistic that can be significantly more sensitive than the SMD in 
certain cases is the {\it Maximum Sample Difference} (MSD). For an one 
dimensional set of data the MSD statistical variable $Max(r_k)$ is defined as 
\be
Max(r_k) = Max({{Y_k}\over {\sigma (Y_k)}})
\ee
where $Y_k$ is given by equations (15), (16) and $\sigma (Y_k))$ is the 
standard deviation of $Y_k$ given by
\be
\sigma(Y_k) = \sqrt{{1\over k} + {1\over {n-k}}}
\ee
The variable $r_k$ has variance unity and mean
\ba
\mu (\alpha, n, k, i_0) &=& 2 \alpha {{(n-i_0)/(n-k)} \over {\sigma (Õ_k)}}  \hspace{1cm} 1<k<i_0 \\
\mu (\alpha, n, k, i_0) &=& 2 \alpha {{i_0/k} \over {\sigma (Õ_k)}}  \hspace{1cm} i_0<k<n-1
\ea

\begin{figure}
\centerline{
\psfig{figure=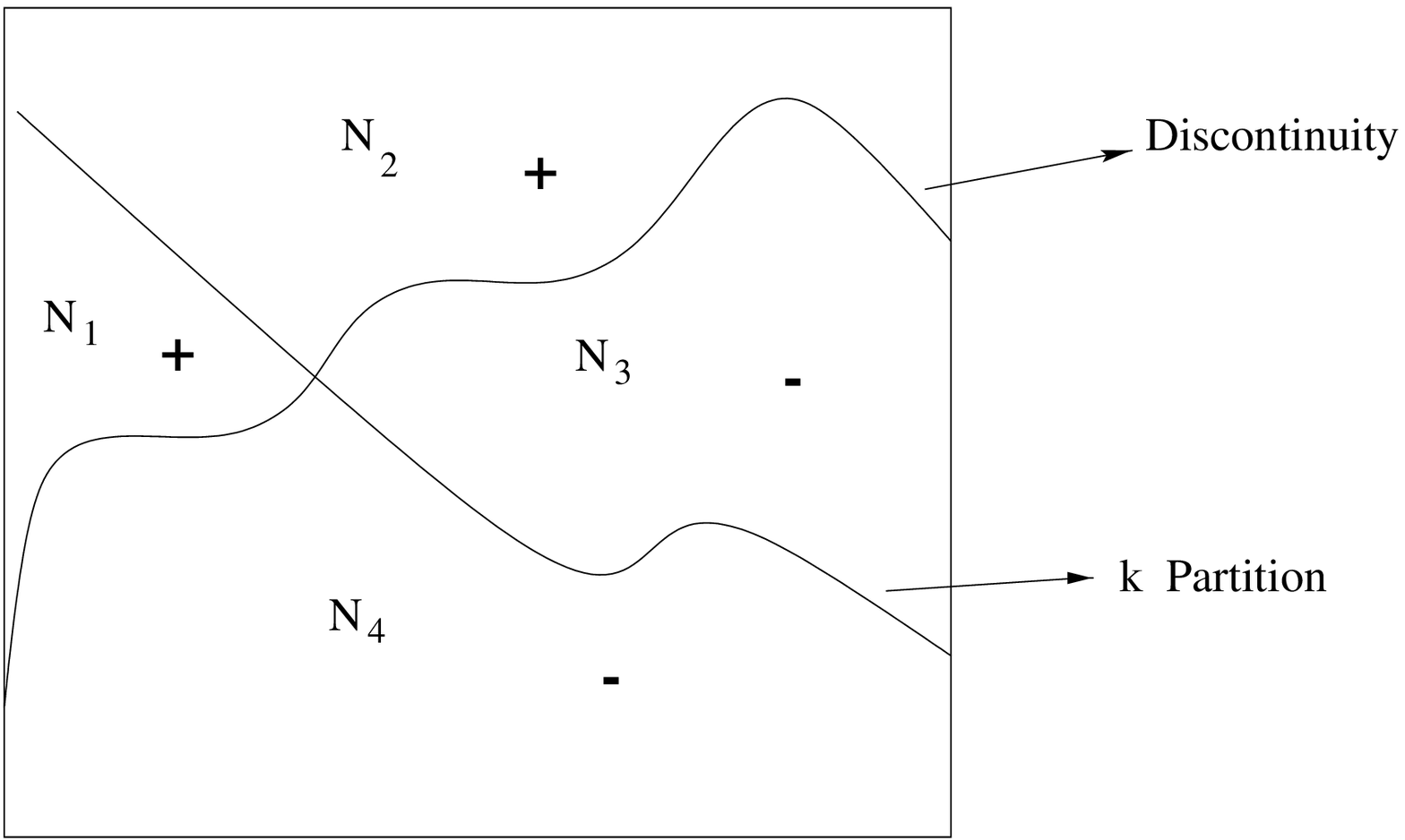,height=2.0in,angle=0}
\psdraft
}Figure 2: Ôhe sample division described by $k$ and the coherent discontinuity 
location described by $i_0$ divide the 2d pixel-map into four parts with 
corresponding number of pixels $N_i$, $i=1,...,4$.
\end{figure}

For two dimensional datasets the index $\bar{k}$ labels {\it partitions} by 
which the two dimensional pixel-surface is divided in two parts. 
In the Monte-Carlo simulations studied here we have considered 
only map divisions represented by straight lines. It is straightforward 
however to generalize this to other types of divisions. In the 2d case $Y_{\bar k}$ 
is generalized to the expression given in equation (38). Let the set of 
parameters $\bar{i_0}$ describe the location and shape of the coherent 
discotinuity in the 2d map (in the simplest case of a straight line 
discontinuity $\bar{i_0}$ represents only two numbers).  Let also 
the set of parameters $\bar{k}$ describe the location of the sample 
division of the 2d map. With the sample division described by $\bar{k}$ and 
the coherent discontinuity location described by $\bar{i_0}$ the 2d pixel-map 
is divided into four parts with corresponding number of pixels $N_i$, $i=1,...,4$ 
(Figure 2). For simplicity, hereafter we will omit the bar in ${\bar{i_0}}$ and 
${\bar{k}}$ thus using the same notation as in the 1d case.

Thus
\ba
Y_{k} &=& {{S_1} \over {N_1 + N_4}} - {{S_2} \over {N_2 + N_3}} \\
\sigma (Y_{k}) &=&  \sqrt{{1 \over {N_1 + N_4}} + {1 \over {N_2 + N_3}}}
\ea
with
\be
r_{k} \equiv {{Y_{k}} \over {\sigma (Y_{k})}} = u_{k}  + 2 \alpha {{w_{k} (i_0)} 
\over {\sigma (Y_{k})}}
\ee
where $u_{k} \equiv {{\Delta \bar{x_{k}}} \over {\sigma(Y_{k})}}$ is a 
standardized Gaussian random variable and 
\be
w_{k}(i_0) = {1 \over 2} ({{N_1 - N_4} \over {N_1 + N_4}} - {{N_2 - N_3} \over {N_2 + N_3}}) 
\ee
Define now $\alpha_{eff} = \alpha w_{k} (i_0)$. Clearly when the partition 
$k$ coincides with the discontinuity $i_0$ ($N_4 \rightarrow 0$ and $N_2 \rightarrow 0$) 
we have $\alpha_{eff} \rightarrow \alpha$. Otherwise $|\alpha_{eff}| < |\alpha | $.
The statistical variable $r_{k}$ is Gaussian with variance unity and mean 
\be
<r_{k}> =  {{2\alpha_{eff}}\over {\sigma(Y_{k})}} \leq {{2\alpha} \over {\sigma(Y_{k})}}
\ee
The $Max (\vert r_{k} \vert)$ after $n$ trials (partitions) is 
therefore a sensitive function of $\alpha_{eff}$ (in the limit where we take {\it all} 
possible partitions we will also have a partition with $\alpha_{eff} \rightarrow \alpha$). 

Now assume that after $n$ trials (partitions) we found $Max (\vert r_{k} \vert) = V_0 > 0$. 
Since the variable $u_k$ of equation (53) is standardized Gaussian the probability 
$p_> (V_0)$ at {\it each trial} that we obtain a value $V_0$ or larger for $\vert r_k \vert $ is 
\ba
p_> (V_0,\alpha_{eff}) &=& {1\over {\sqrt{2\pi}}} \int_{\vert V_0 \vert}^\infty dr_{k} 
e^{-{{(r_{k} - (2 \alpha_{eff}) /\sigma(Y_{k})))^2} \over 2}} \nonumber \\
&+&  
{1\over {\sqrt{2\pi}}} \int_{-\infty}^{-\vert V_0 \vert} dr_{k} 
e^{-{{(r_{k} - (2 \alpha_{eff} /\sigma(Y_{k})))^2} \over 2}} 
\ea
Using the binomial distribution we find the probability for $x$ values of $r_k$ 
above $V_0$ after $n$ partitions to be
\ba
P_x (n,V_0,\alpha_{eff}) &=& \nonumber \\
{{n!}\over {x! (n-x)!}} p_> (&V_0&, \alpha_{eff})^x (1-p_> 
(V_0,\alpha_{eff}))^{n-1}
\ea
In our case we have {\it only one} occurence of $V_0$ (since it is maximum) and the 
probability for this to happen is $P_1 (n,V_0,\alpha_{eff})$. Thus, from a 2d 
pixel-map we can measure $V_0$ (the maximum of $r_{k}$, $n$ (the number of divisions 
used in the test) and $\sigma(Y_{{k_0}})$ (for the partition $k_0$ that corresponds to $V_0$)). 
With this input we obtain the probability distribution $P_1 (\alpha_{eff})$ given $n$, 
$V_0$ and $\sigma (Y_{{k_0}})$. For example assume that we measured $V_0$ with 100 
divisions ($n=100$) in a $30 \times 30$ pixel map. A reasonable value of  
$\sigma (Y_{{k_0}})$ (to be obtained exactly from the data) is 
\be
\sigma (Y_{\bar{k_0}}) = \sqrt{{1 \over {N_1 + N_4}} - 
{1 \over {N_2 + N_3}}}\simeq {2 \over \sqrt{N}} \simeq 0.07
\ee
Given $n$, $V_0$ and $\sigma (Y_{{k_0}})$, the probability distribution 
$P_1 (\alpha_{eff})$ is an even function of $\alpha_{eff}$, completely 
determined and has maxima $P_1(\alpha_{eff}^{max})$ at $\pm \alpha_{eff}^{max}$. 
For larger $V_0$ we expect larger $\alpha_{eff}^{max}$.

For example, it may be easily verified using equation (57) and the package M
athematica that with $n=100$ we have $|\alpha_{eff}^{max}| = 0$ for 
$V \leq V_0^{crit} \simeq 2.5$. In general, given $n$, $V_0$ and 
$\sigma (Y_{\bar{k_0}})$, we can determine the probability distribution for 
$\alpha_{eff}$ and therefore the most probable value of $\alpha_{eff}$ 
(a lower bound on $\alpha$) can be found. This most probable value is 
$\alpha_{eff}^{max}$ and the corresponding probability is $P_1 (\alpha_{eff}^{max})$. 
We also find the probability that there is no coherent discontinuity on 
the map as $P_1 (\alpha_{eff} = 0)$ (for $V_0 < V_0^{crit}$ it is most 
probable that there is no coherent discontinuity on the map).

\begin{figure}
\centerline{
\psfig{figure=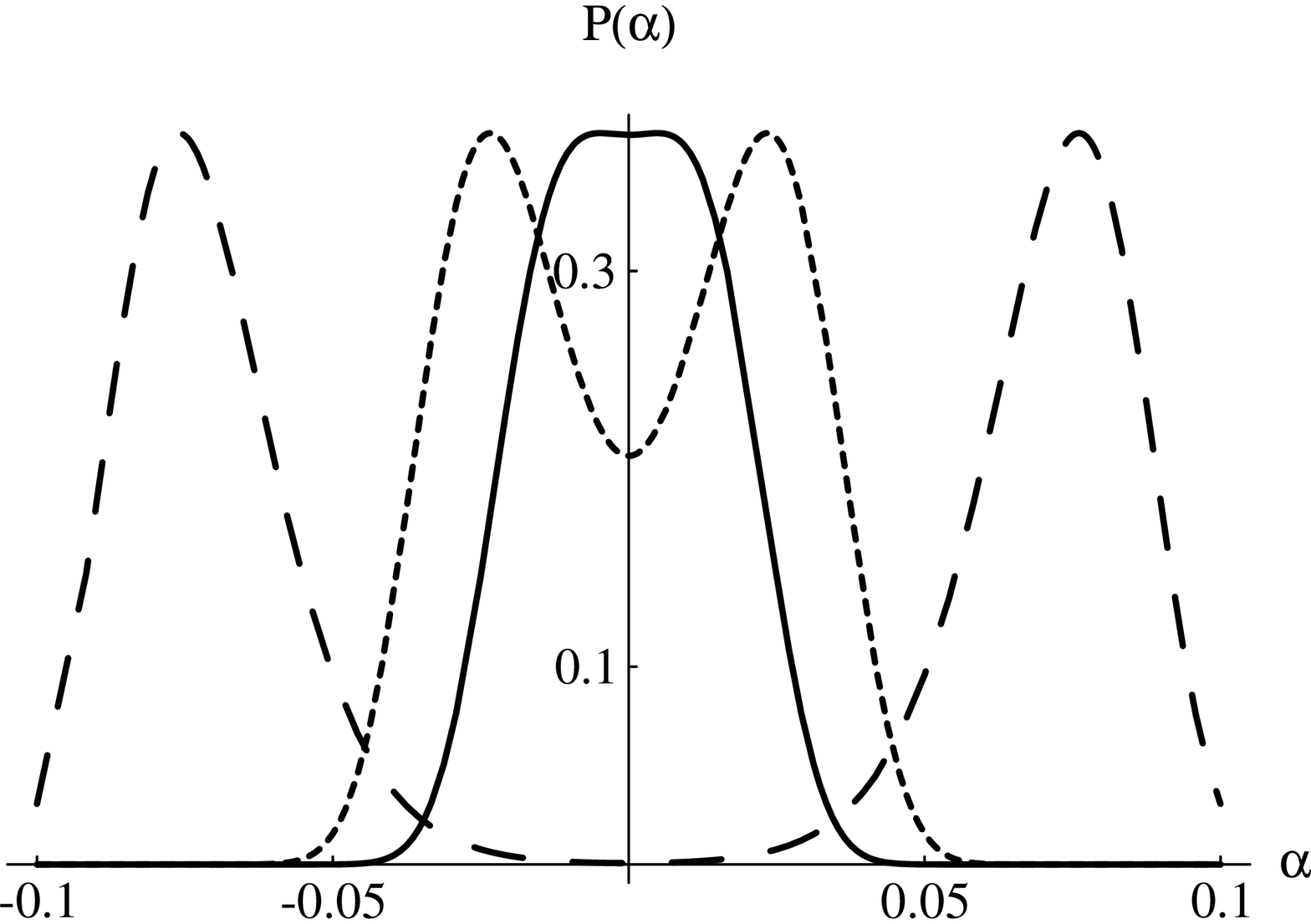,height=2.4in,angle=0}
\psdraft
}
Figure 3: Á plot of $P_1(\alpha_{eff})$ for $V_0 = 2.6$ (continous line), $V_0 = 3.0$  
(dotted line) and $V_0 = 4.5$ (dashed line).
\end{figure}

In Figure 3 we show a plot of $P_1 (\alpha_{eff})$ for $V_0 = 2.6$ 
(continous line), $V_0 = 3.0$ (dotted line) and $V_0 = 4.5$ (dashed line). 
Clearly for $V_0 \leq 2.6$ it is most probable that there is no coherent 
discontinuity on the map ($\alpha_{eff}^{max} = 0$) with $P_1 (\alpha_{eff} = 0) = 0.33$. 
On the other hand for $V_0 = 3.0$, the most probable value of $\alpha_{eff}$ is 
$|\alpha_{eff}^{max}| = 0.04$ with probability $P_1(\alpha_{eff}^{max}) \simeq 0.37$ 
while the probability that there is no coherent discontinuity on the map 
is $P_1 (\alpha_{eff} = 0) \simeq 0.16$.

It is important to verify the above analytical results using Monte Carlo 
simulations of 2d data. We considered 2d $30 \times 30$ data-sets as described in 
section 3 with uncorrelated standardized Gaussian data (white noise). On these we 
superpose a coherent discontinuity with amplitude $2\alpha$ with $\alpha$ in the 
range $[0, 0.45]$. For each $\alpha$ we construct 10 maps and find $V_0$ and its 
standard deviation $\sigma (V_0)$. We also find $\sigma (Y_{\bar{k_0}})$ which was 
practically constant at  $\sigma (Y_{\bar{k_0}}) \simeq 0.07$ as predicted analytically. 
With the input $V_0$ and $\sigma (Y_{\bar{k_0}})$ we construct $P_1 (\alpha_{eff})$ 
and find $|\alpha_{eff}^{max}|$, $P_1 (\alpha_{eff}^{max})$ and $P_1 (0)$. 
These results are shown in Table 5.

{\bf Table 5}:  Ôhe effectiveness of the MSD statistic considered in 
two dimensional maps with a flat spectrum of fluctuations (white noise). 
First column shows the magnitude of the coherent discontinuity superposed 
on the standardized Gaussian map and the fourth column shows shows the 
{\it derived} most probable value value of $\alpha_{eff}$ (the lower bound 
of $\alpha$) based on the MSD statistic.
\vskip 0.1cm
\begin{tabular}{|c|c|c|c|c|c|}\hline
{\bf $\alpha$ }&{\bf  $V_0$} &{ \bf  $\sigma (Y_{\bar{k_0}})$ }
& {\bf $|\alpha_{eff}^{max}|$}& {\bf $P_1(0)$}   & {\bf $P_1(|\alpha_{eff}^{max}|)$} \\ \hline
0.00 &$2.6 \pm 0.5$     & $0.07  $     &$0.0 \pm 0.01$  &$0.36 $    &  $0.36 $      \\ \hline
0.03 &$2.8 \pm 0.6$     & $0.07  $     &$0.02 \pm 0.02$  &$0.31 $   &  $0.37 $      \\ \hline
0.06 &$3.4 \pm 0.7$     & $0.07  $     &$0.04 \pm 0.02$  &$0.06 $   &  $0.37 $      \\ \hline
0.1  &$4.1 \pm 0.7$     & $0.07  $     &$0.06 \pm 0.03$  &$0.005 $  &  $0.37 $      \\ \hline
0.2  &$5.6 \pm 0.7$     & $0.07  $     &$0.11 \pm 0.03$  &$0.0 $    &  $0.37 $      \\ \hline
0.25  &$7.5 \pm 0.7$    & $0.07  $    &$0.18 \pm 0.03$  &$0.0 $    &  $0.37 $      \\ \hline
0.3  &$9.0 \pm 0.9$     & $0.07  $     &$0.24 \pm 0.03$  &$0.0 $    &  $0.37 $      \\ \hline
0.35  &$9.8 \pm 0.8$    & $0.07  $    &$0.27 \pm 0.03$  &$0.0 $    &  $0.37 $      \\ \hline
0.4  &$10.4 \pm 1.1$     & $0.07  $    &$0.29 \pm 0.03$  &$0.0 $    &  $0.37 $      \\ \hline
0.45  &$12.4 \pm 1.6$    & $0.07  $    &$0.36 \pm 0.04$  &$0.0 $    &  $0.37 $      \\ \hline
\end{tabular}

Comparing $\alpha_{eff}$ with $\alpha$ we confirm that in all simulated cases 
$|\alpha_{eff}^{max}|$ is a lower bound on $|\alpha |$. It is also clear from 
Table 5 that the MSD method can detect the presence of a coherent discontinuity 
with $\alpha \gsim 0.04$ with probability ratio 
\be
{{P_1 (\alpha_{eff} = 0.04)} \over  {P_1 (\alpha_{eff} = 0)}} \simeq 7
\ee
The MSD statistic is significantly more sensitive in detecting coherent 
discontinuities compared to the SMD statistic of sections 2, 3. 
We expect however that the sensitivity of this statistic will be significantly 
reduced when correlations are introduced in the data. We have shown using additional 
Monte-Carlo simulations that this statistic is not as robust as the SMD statistic. 
In particular when we include correlations in the data (\eg scale invariance), the 
sensitivity of the MSD statistic drops rapidly to the level of the SMD statistic \ie  
it can detect a coherent discontinuity with $\alpha \gsim 0.4$. This implies that 
the MSD statistic is more useful in detecting the presence of coherent 
discontinuities only when applied to noise dominated data.

\section{\bf Concluding Remarks}

An important issue that needs to be clarified is the following:'What are the effects 
of other strings giving rise to their own step discontinuity? Do they decrease the 
sensitivity of the suggested statistical tests?'

No attempt is made in this paper to model the fluctuations of 'other strings'. Any 
such attempt (even those of simulations) is faced with the possibility of serious 
errors due to incorrect assumptions. Even basic features of the string scaling solution 
are still under serious debate. For example there have been serious claims recently 
\cite{vhs97} that realistic field 
theoretical cosmological simulations of gauged string evolution would have no 
wiggles for long strings and no loop component. In addition, the physical processes 
affecting the CMB photons are not well known especially in defect based models. 
The issues of reionization, fluctuations present on the last scattering surface, 
wiggles of long strings and other effects have only been crudely modeled so far. 

Instead of attempting a rough modeling of these effects we have made a very robust 
and reasonable assumption: The statistics of CMB fluctuations induced by a string 
network on large angular scales are either gaussian (as was the common belief so far) 
or minimally non-gaussian in the sense that the only non-gaussianity is due to a 
late long string. Additional types of non-gaussianity are not excluded but they would 
simply make the detection of non-gaussianity easier by using the proposed (or other) 
tests. In that sense the proposed tests would only be able to find a lower bound on 
$G\mu$ which however turns out to be cosmologically quite interesting given the 
optimum sensitivity of the tests for detecting the above defined 'minimal non-gaussianity'.

The question that has been addressed in this paper is the following: 
Given the presently known values for ${{\delta T} \over T}_{rms}$ from COBE on large 
angular scales, what is the minimum value of $G\mu$ detectable under the above stated 
assumption of 'minimal non-gaussianity'?
Using the SMD or MSD statistics which are optimized to detect coherent temperature 
discontinuities on top of Gaussian temperature maps we may obtain non-trivial 
upper or even {\it lower} bounds on the values of $G\mu v_s \gamma_s$ which are 
highly robust and independent of the details of the string evolution and the 
resolution of the CMB maps. Application of these statistics on the COBE data is 
currently in progress \cite{ps97}.

\acknowledgments

I wish to thank N. Simatos and T. Tomaras for
interesting discussions and for providing helpful comments after reading the 
paper. 
This work is the result of a network supported by the European Science
Foundation. 
The European Science Foundation acts as catalyst for the development of 
science by bringing together leading scientists and funding agencies to
debate, plan and implement pan-European initiatives.
This work was also supported by the EU grant CHRX-CT94-0621 as well 
as by the Greek General Secretariat of Research and Technology grant
$\Pi$ENE$\Delta$95-1759.

\vfill
                      
\eject

\end{document}